\pdfoutput=1
\documentclass[10pt]{article}

\usepackage[T1]{fontenc}
\usepackage{lmodern}
\usepackage{amsmath}
\usepackage{booktabs}
\usepackage{array}
\usepackage{graphicx}
\usepackage[hypertexnames=false]{hyperref}
\usepackage[expansion=false]{microtype}
\usepackage{tikz}
\usepackage{pgfplots}
\usepackage{tcolorbox}
\usepackage{float}
\usepackage{placeins}
\usepackage{needspace}
\usepackage{xurl}
\usepackage{pdflscape}

\usetikzlibrary{arrows.meta,positioning,fit,backgrounds}
\pgfplotsset{compat=1.18}
\tcbuselibrary{listings,skins,breakable}

\newtcblisting{codebox}{
  listing only,
  breakable,
  colback=gray!5,
  colframe=gray!60,
  boxrule=0.4pt,
  left=6pt,right=6pt,top=4pt,bottom=4pt,
  listing options={basicstyle=\ttfamily\footnotesize,breaklines=true,columns=fullflexible}
}

\newcommand{\DatasetTinder}{\textsc{Tinder (Google Play)}}
\newcommand{\DatasetGojek}{\textsc{Gojek (Google Play)}}
\newcommand{\DatasetTwitter}{\textsc{Customer Support on Twitter}}
\newcommand{\DatasetAmazon}{\textsc{Amazon Appstore (Low-rated)}}
\newcommand{\DatasetAmazonAnn}{\textsc{Amazon Appstore (Annotated)}}

\newcommand{\MethodLex}{\textsc{Lexical}}
\newcommand{\MethodTFIDF}{\textsc{TF-IDF}}
\newcommand{\MethodEmbed}{\textsc{Embedding}}

\newcommand{\UXResolve}[2]{\def#1{x}}

\title{Proxy-Validated LLM UX Micro-Simulations: An Artifact-First Protocol for Early-Stage Decision Support}
\author{Alexandre Cristov\~ao Maiorano\\alexandre@lumytics.com}
\date{}

\begin{document}
% Allow LaTeX a small emergency stretch to break lines that contain long
% unbreakable tokens (\texttt with underscores, run IDs); avoids overfull hboxes
% without resorting to \sloppy globally.
\emergencystretch=2em
\maketitle

\begin{abstract}
Early-stage teams often lack users, time, and budget to run repeated UX studies,
yet still need decision-oriented signals to iterate safely.
We study an LLM-driven \emph{UX micro-simulation} pipeline that generates structured
customer-experience feedback (walkthrough steps, friction points, micro-survey signals)
from versioned prompts, personas, tasks, and UI snapshots.
Because public usability datasets with task outcomes are scarce, we validate simulated
friction themes using multiple public \emph{proxy} corpora (app reviews, support tweets,
and open-source software issues). We propose a lightweight proxy-validation protocol with two alignment metrics---top-$k$ Jaccard and distributional weighted-Jaccard ($W$)---and compare lexical, TF-IDF, and multilingual embedding baselines across six proxy datasets.
Embedding-based alignment yields higher $W$ than lexical baselines on primary app-review and support-tweet proxies (e.g., $W{=}0.128$ vs $0.000$ on Gojek), while top-$k$ Jaccard is shown to overstate alignment at large $k$.
We ablate four agent strategies (single-pass, best-of-$N$, hybrid, and a proposed \emph{score-then-select} judge) across Azure OpenAI deployments and report bootstrap confidence intervals over $8$ method--dataset pairs; these intervals reveal that the embedding $W$ point estimate is systematically unstable under resampling at our subsample size.
We also provide a failure-mode analysis of grounding and fabrication proxies, with documented calibration caveats and worked examples of outputs flagged as fabricated by an adversarial judge.
Our artifact-first pipeline produces reproducible tables and figures from versioned run
artifacts, supporting iterative prompt and taxonomy refinement before final paid-model
calibration.
\end{abstract}

\section{Introduction}
UX evaluation is costly, slow, and often deferred until a product has enough users to justify studies.
For early-stage teams, this creates a gap: product decisions must be made under uncertainty,
while regressions and UX friction can accumulate silently until they become expensive to unwind.
In parallel, large language model (LLM) tooling has made it feasible to generate structured, role-conditioned feedback at scale,
but the scientific and practical question remains: \emph{what does such synthetic feedback mean}, and how can
we validate it without over-claiming?

This paper studies a pragmatic middle ground: we use LLM-driven \emph{micro-simulations} to generate
structured UX feedback from versioned prompts, and validate the resulting friction signals against
multiple public \emph{proxy datasets} (app reviews, customer support tweets, open-source software issues).
Rather than treating proxies as ground truth for task success, we treat them as noisy distributions of
friction themes that can be compared to simulation outputs under controlled configurations.

\paragraph{Research questions and hypotheses.}
We focus on three practical questions:
(RQ1) do simulation outputs align with friction distributions observed in public proxy corpora?
(RQ2) which proxy-alignment method is most robust (lexical vs multilingual embeddings, and which overlap metric)?
and (RQ3) are conclusions stable under sampling variability and prompt/taxonomy changes?
Our main hypotheses are that embedding-based overlap will outperform lexical baselines under weighted-Jaccard (H1),
that top-$k$ Jaccard saturates and can overstate alignment when $k$ grows (H2),
and that aggregate alignment estimates are sufficiently stable to support iterative decision-making when reported with bootstrap intervals (H3).

\paragraph{Contributions.}
We make three contributions:
(1) a reproducible UX micro-simulation pipeline with structured outputs and versioned inputs;
(2) a proxy-validation protocol with alignment metrics (top-$k$ and weighted Jaccard),
including lexical and multilingual embedding baselines; and
(3) an artifact-first evaluation suite with bootstrap confidence intervals and error analyses, designed
to be upgraded with stronger paid models only at the end of the iteration loop.

\paragraph{Value to practitioners.}
For product and engineering teams, the output of our pipeline is not a single ``UX score,'' but a set of actionable signals:
(i) a ranked set of friction themes to prioritize in the backlog, (ii) a compact summary of where the current experience
is likely to fail (by locale/domain proxy), and (iii) an auditable, versioned trail of prompts, taxonomies, and artifacts
that makes changes explainable and reviewable.
In other words, we turn otherwise unstructured, costly-to-collect ``user feedback'' into a reproducible iteration loop
that can be consumed by stakeholders who need to decide \emph{what to fix next} and \emph{what risks are acceptable} in early stages.

\section{Background and Related Work}
\paragraph{Inspection-based UX evaluation.}
Cognitive walkthroughs are a classic, theory-driven inspection method for identifying usability issues without running user studies~\cite{polson1992cognitive}.
At scale, organizations adopt metrics frameworks such as HEART to connect product goals to measurable signals and metrics~\cite{rodden2010heart}.
The System Usability Scale (SUS) remains a widely used, lightweight instrument for perceived usability~\cite{brooke1996sus}.

\paragraph{Mining user feedback at scale.}
App reviews and related feedback sources have been extensively studied in requirements engineering and software maintenance,
with systematic surveys cataloging use cases and evaluation gaps~\cite{dabrowski2022review}.
Public corpora of reviews, support conversations, and issues enable reproducible studies, but they rarely include
task completion times or controlled experimental conditions.
Recent datasets focus on structured taxonomies of end-user issues, including low-rated Amazon Appstore reviews with a labeled subset~\cite{amazonlowrated2026dataset}.

\paragraph{LLMs as simulators and the ``human surrogate'' debate.}
Recent work explores ``silicon samples'' and LLM-based simulation of human responses~\cite{argyle2023outofone,aher2023simulate,horton2023homosilicus}.
In HCI, LLM-powered agents and personas have been proposed for scalable simulation and ideation~\cite{park2023generativeagents,choi2025proxona,schuller2024personas,prpa2024syntheticpersonae}.
However, multiple studies highlight validity risks and misalignment between LLM outputs and human distributions~\cite{bisbee2024perils,amirova2024algorithmicfidelity,gao2024scylla,li2025personacatch}.
We position our work as \emph{decision support for early iteration}, not a replacement for human studies:
we use proxy validation to detect egregious mismatches and to guide prompt/taxonomy iteration,
while reserving stronger paid-model calibration for the final evaluation pass.

\paragraph{LLM-based evaluation and rubric reliability.}
LLM-as-a-judge methods are increasingly used as scalable approximations to human evaluation~\cite{liu2023geval,zheng2023mtbench},
but they can exhibit systematic biases (e.g., position and verbosity effects) and inconsistent rubric application~\cite{wang2024notfair,shi2024positionbias,stureborg2024inconsistent}.
We therefore emphasize (i) schema-constrained outputs, (ii) decision trails tied to concrete artifacts, and (iii) small gold-labeled calibration sets when available.

\paragraph{Positioning.}
Unlike prior LLM simulation work that targets individual-level behavioral fidelity or post-hoc user-study surrogates, our contribution is a \emph{protocol} layer: a lightweight proxy-validation loop that quantifies distributional alignment between simulation outputs and public corpora, with reproducible alignment metrics, bootstrap confidence intervals, and artifact-first reporting.
This complements existing frameworks by providing structured go/no-go signals for prompt and taxonomy iteration \emph{before} committing to expensive human studies or large-scale model calibration.

\section{Method}
\subsection{UX micro-simulations}
Our pipeline generates structured feedback by sampling a persona, journey, and task, then rendering a UI-context
snapshot (labels, metrics, and a scenario outcome) and prompting an LLM to answer as the persona.
Each simulation emits JSON fields including:
(i) step-by-step walkthrough, (ii) friction points mapped to a fixed taxonomy,
(iii) sentiment and suggested fixes, and (iv) micro-survey signals---customer satisfaction (CSAT), customer effort score (CES), and net promoter score (NPS)---and SUS-like answers.
We enforce a strict structured-output contract (schema + lightweight prompt rules) to keep outputs comparable across prompt versions and agent strategies.
The pipeline carries two simulator prompt versions: \texttt{v1} (baseline) and \texttt{v2} (which adds explicit grounding rules and a friction-taxonomy prefix); all experiments in this paper use \texttt{v2}, and we keep \texttt{v1} only as the prior baseline for context (excerpts in Appendix).
All prompts and input configs are versioned to ensure reproducibility.
Figure~\ref{fig:pipeline} sketches the end-to-end workflow: versioned inputs (prompts, personas, journeys, UI snapshots, taxonomy) feed the simulator; structured outputs are scored against multiple proxy corpora via the alignment metrics defined below; tables and figures are exported directly from run artifacts.
\UXResolve{\uxpipelinefig}{figures/pipeline_overview.tex}
\ifx\uxpipelinefig\empty\else
\begin{figure}[t]
  \centering
  \resizebox{0.98\linewidth}{!}{\begin{tikzpicture}[
  font=\small,
  box/.style={draw, rounded corners, align=center, minimum height=14mm, minimum width=30mm},
  arrow/.style={-{Latex[length=2mm]}, thick, shorten <=2pt, shorten >=2pt},
  group/.style={draw=gray!60, rounded corners, inner sep=6pt}
]
\node[box] (inputs) {Versioned inputs\\\footnotesize prompts, personas, tasks,\\taxonomy, UI snapshot};
\node[box, right=18mm of inputs] (llm) {LLM micro-simulation\\\footnotesize schema-constrained JSON};
\node[box, right=18mm of llm] (agg) {Aggregate signals\\\footnotesize theme counts,\\micro-surveys};

\node[box, below=14mm of llm] (proxies) {Proxy corpora\\\footnotesize reviews, support,\\OSS issues};
\node[box, right=18mm of proxies] (align) {Alignment\\\footnotesize $W$, $J_k$, bootstrap CI};

\node[box, right=18mm of agg] (outputs) {Decision artifacts\\\footnotesize backlog + summary\\CSV/JSON/\LaTeX};

\begin{scope}[on background layer]
  \node[group, fit=(inputs) (llm) (agg) (outputs)] (topgroup) {};
  \node[group, fit=(proxies) (align)] (bottomgroup) {};
\end{scope}

\draw[arrow] (inputs.east) -- (llm.west);
\draw[arrow] (llm.east) -- (agg.west);
\draw[arrow] (agg.east) -- (outputs.west);
\draw[arrow] (proxies.east) -- (align.west);
\draw[arrow] (agg.south) -- (align.north);
% Route around the aggregate box (avoid overlapping).
\draw[arrow] (align.east) -| (outputs.south);
\end{tikzpicture}}
  \caption{Micro-simulation and proxy-validation workflow (artifact-first).}
  \label{fig:pipeline}
\end{figure}
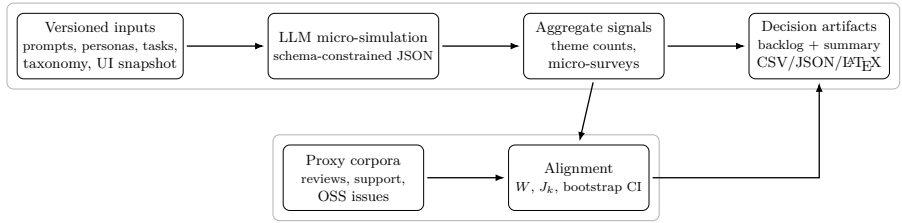
\fi

\subsection{Agent strategies (single-pass, best-of-\texorpdfstring{$N$}{N}, hybrid)}
\label{sec:agent-strategies}
We support three practical strategies for producing schema-compliant simulation outputs under a fixed prompt and model:
(i) \emph{single-pass} (one attempt with bounded retries),
(ii) \emph{best-of-$N$} (sample $N$ candidates, then select one via an LLM judge), and
(iii) \emph{hybrid} (attempt single-pass first; fall back to best-of-$N$ only on validation failure).
We additionally include a fourth strategy, \emph{score-then-select} (Section~\ref{sec:anchoring-aware-judge}), which scores candidates independently before picking the best one.
We use these four names consistently throughout the paper, including the table columns.
We report quality--cost tradeoffs from the agent ablation experiments in Section~\ref{sec:anchoring-aware-experiment} (Tables~\ref{tab:agent-ablation-hq} and \ref{tab:agent-ablation-cq}).

\subsection{Proposed: score-then-select judge}
\label{sec:anchoring-aware-judge}
\textbf{Motivation.} Both human and LLM-mediated evaluation pipelines can exhibit anchoring and order effects, where early impressions disproportionately shape later judgments~\cite{tversky1974judgment,shi2024positionbias}. For example, AgentReview discusses anchoring as a plausible mechanism for why later evidence may not overturn initial ratings in multi-agent peer review settings~\cite{jin2024agentreview}. This motivates a stricter separation between \emph{generation} and \emph{selection} in best-of-$N$ settings.

\paragraph{Method.}
Given $N$ candidate simulation outputs (same inputs, different sampling), we score candidates \emph{independently} using a fixed rubric, without showing the judge multiple candidates in the same prompt. For each candidate $i$, the judge returns a small set of scalar rubric scores (e.g., schema compliance, UI-grounding, and clarity), plus a short justification. We then select the winning candidate deterministically (e.g., by weighted sum of rubric scores), and optionally run a tie-break on the top-$2$ candidates with randomized order and a win-by-both-orders rule to reduce position bias~\cite{zheng2023mtbench,shi2024positionbias}. This ``score-then-select'' design is intended to reduce anchoring compared to direct pairwise comparison or multi-candidate ranking in a single context window~\cite{tversky1974judgment}, echoing the concern raised in AgentReview that early reviewer impressions can dominate later evidence and anchor decisions~\cite{jin2024agentreview}. In a cost-sensitive setting, this can be combined with our hybrid gating: only invoke candidate generation and judging when single-pass fails validation or falls below a threshold on cheap heuristics (e.g., prefix and grounding proxies).

\paragraph{Selection protocol (what the judge sees).}
In best-of-$N$, the judge receives all candidates together (with randomized order) and returns the index of the one it picks. In score-then-select, the judge sees one candidate at a time, returns rubric scores, and selection is $\arg\max_i(\text{overall}_i)$.
\begin{quote}\footnotesize\ttfamily
best-of-N: candidates\_json=[cand\_a,cand\_b] $\rightarrow$ selected\_index \\
score-then-select: candidate\_json=cand\_i $\rightarrow$ overall\_i; select argmax
\end{quote}
This makes the selection protocol explicit and reduces ambiguity about potential cross-candidate anchoring.

\subsection{Friction taxonomy and domain weighting}
We use a fixed friction taxonomy (e.g., latency, errors, navigation, configuration, traceability) with optional domain-specific weights.
When validating against proxy datasets, we apply simple domain stopwords and weights to reduce spurious keyword matches.
In practice, this means (i) filtering common non-informative tokens (e.g., generic ``app'' terms in reviews) and
(ii) optionally up-weighting categories that are expected to dominate a proxy (e.g., reliability and latency for observability tools),
while keeping the taxonomy and weighting rules versioned and explicit.

\subsection{Proxy validation and alignment metrics}
Because public datasets rarely provide task success, we evaluate \emph{theme alignment} between simulation frictions
and proxy frictions computed from text corpora.
Given simulated friction counts $S$ and proxy friction counts $P$ over a shared taxonomy:

\paragraph{Top-$k$ Jaccard.}
We take the top-$k$ categories of each distribution, $Top_k(S)$ and $Top_k(P)$, and compute:
\[
J_k = \frac{|Top_k(S) \cap Top_k(P)|}{|Top_k(S) \cup Top_k(P)|}.
\]
We treat $J_k$ as a coarse, human-auditable signal; when $k$ is large relative to the taxonomy size, $J_k$ can become artificially high.

\paragraph{Weighted-Jaccard.}
We compute a distributional overlap:
\[
W = \frac{\sum_i \min(S_i, P_i)}{\sum_i \max(S_i, P_i)}.
\]
We interpret $W \in [0, 1]$ as the fraction of shared ``mass'' between two discrete distributions, where 1.0 indicates identical category distributions.

\paragraph{Alignment methods.}
We compare:
\MethodLex: keyword and phrase matching;
\MethodTFIDF: a similarity baseline based on term frequency--inverse document frequency (a standard weighting that downplays common words) for assigning each text to a single best-matching category if above a threshold; and
\MethodEmbed: multilingual embedding similarity using a local embedding model (BGE-M3 via Ollama) with a similarity threshold.
To keep local compute feasible, embedding alignment optionally uses a fixed subsample size (we use $n=200$ per corpus unless noted).

\subsection{Uncertainty estimation}
We report bootstrap confidence intervals---ranges obtained by repeatedly resampling the data with replacement---for alignment metrics, to quantify how much each metric would vary under sampling.

\subsection{Assumptions and claims (scope)}
We make three bounded claims:
(C1) micro-simulations can produce structured friction signals that are useful for early-stage iteration;
(C2) proxy alignment can detect \emph{egregious} mismatches between simulated and observed friction distributions; and
(C3) embedding-based alignment tends to better match our taxonomy than purely lexical baselines under the same protocol \emph{in single-shot point estimates}, especially on our primary proxies; we treat the bootstrap stability of the embedding $W$ statistic as a separate, empirically tested question (Section~\ref{sec:results-bootstrap}).

We explicitly do \emph{not} claim that simulated outputs match human behavior at the individual level, nor that proxy alignment implies task success or user preference.

\section{Datasets and Proxies}
We validate simulations against multiple public proxy datasets; the record counts below are those loaded in our experiments:
\begin{itemize}
  \item \DatasetTinder~\cite{kaggle_tinder}: English app reviews with ratings ($\approx 664$k records).
  \item \DatasetGojek~\cite{kaggle_gojek}: Indonesian app reviews with ratings ($\approx 1.67$M records).
  \item \DatasetTwitter~\cite{kaggle_twitter_support}: customer support tweets ($\approx 2.81$M records; no star ratings).
  \item Open-source software (OSS) issues proxy: GitHub issues from business-to-business (B2B) observability tools (Grafana, Prometheus, OpenTelemetry Collector, OpenTelemetry JS, Jaeger, Zipkin, Sentry), sampled for feasibility.
  \item \DatasetAmazon~\cite{amazonlowrated2026dataset}: low-rated Amazon Appstore reviews (79{,}531 reviews after dropping empty bodies).
  \item \DatasetAmazonAnn~\cite{amazonlowrated2026dataset}: labeled subset with issue-type categories (8{,}971 annotated reviews).
\end{itemize}

We treat these datasets as \emph{proxies} for friction themes rather than as ground truth for task outcomes.
For the annotated Amazon subset, we additionally perform an LLM-based label evaluation (Section~\ref{sec:appstore-eval}) where the dataset's category serves as a noisy gold label for classification.

For compute feasibility, some alignment variants operate on capped slices of large proxy corpora. Concretely, embedding-based alignment on \DatasetTinder{} and \DatasetGojek{} was run on a pre-cap of $n=10{,}000$ texts, while \DatasetTwitter{} and \DatasetAmazon{} were processed without the pre-cap (the BGE-M3 subsample of $n=200$ per corpus controls embedding cost in all cases).
In all experiments, we keep locale and domain explicit (e.g., app reviews vs support conversations) to avoid overclaiming transfer.

\section{Experiments}
\subsection{Agent ablation (single-pass vs best-of-\texorpdfstring{$N$}{N} vs hybrid)}
Following the strategies defined in Section~\ref{sec:agent-strategies}, we operationalize the research-to-product tradeoff between output quality and inference cost by ablating---that is, systematically comparing under matched conditions---three agent strategies under the same schema constraints:
(i) single-pass (one attempt with retries), (ii) best-of-$N$ (sample $N$ candidates and select one with an LLM judge), and (iii) hybrid (try single-pass first; fall back to best-of-$N$ only when validation fails).
We additionally enforce a lightweight taxonomy contract in the \texttt{v2} prompt: the single friction point must be prefixed with a taxonomy identifier (e.g., \texttt{navigation: ...}).
For the current paper runs, we execute the reduced ablation suites on Azure using OpenAI deployments (\texttt{gpt-4.1}, \texttt{gpt-4.1-mini}, \texttt{gpt-5.2}) with $n=60$ per condition/model. We standardize multi-candidate methods to $N=2$ as the default comparison point and include targeted ablations that stress selection behavior and the cost/quality frontier (Table~\ref{tab:agent-ablation-conditions}). All runs in this paper use the \texttt{v2} simulator prompt; the selection strategy (LLM judge vs.\ judge-free heuristic) is reported as the \emph{Judge} column in Tables~\ref{tab:agent-ablation-hq} and~\ref{tab:agent-ablation-cq}.
To reduce confounds across agent variants, we fix the sampling seed (default \texttt{seed=42}) so each run draws the same persona/journey/task sequence, and we keep model and decoding parameters matched across methods.
For best-of-$N$, we shuffle candidate order~\cite{shi2024positionbias} before presenting them to the judge to reduce position bias.
Under this setting, hybrid remains the practical default: it preserves structured-output compliance while reducing average LLM calls per simulation because many samples are solved in the initial single-pass attempt (Tables~\ref{tab:agent-ablation-hq} and \ref{tab:agent-ablation-cq}); Table~\ref{tab:agent-output-examples} shows representative simulator outputs for the same (persona, journey, task) under different agent strategies.
We also evaluate a judge-free selection mode (the \emph{Auto} judge in the ablation tables) that selects among $N$ candidates using only cheap heuristics (UI token-overlap grounding, task success, and a simple consensus boost over the friction category prefix).
We additionally support stress tests with larger candidate pools and higher candidate temperatures (more diverse generations), but keep the paper core at $N=2$ for comparability and API-budget control.
Best-of-$N$ selection is conceptually aligned with best-of-$N$ sampling and LLM-judge paradigms used in scalable evaluation and selection settings, though such judges can exhibit systematic biases (e.g., position/verbosity)~\cite{wang2023selfconsistency,liu2023geval,zheng2023mtbench,shi2024positionbias}.
\paragraph{Metrics.}
We summarize agent tradeoffs using cheap, locally computed proxies:
\emph{Calls} is the mean number of LLM calls per simulation (including judge calls when applicable), used as a cost proxy.
\emph{Prefix ok} is the rate that the single friction point begins with a taxonomy identifier prefix.
\emph{Quality} is a composite proxy over rates in $[0,1]$:
$0.35\,r_{words}+0.20\,r_{steps}+0.15\,r_{HEART}+0.10\,r_{success}+0.10\,(\overline{SUS}/100)+0.10\,r_{prefix}$.
Weights were chosen to emphasize content completeness ($r_{words}$) and structural compliance ($r_{steps}$, $r_{HEART}$).
$r_{words}$ is 1 if all string fields (feedback, suggested fix, walkthrough steps, HEART items) contain $\leq 12$ words; this is stricter than the validation gate, which only enforces the word limit on \texttt{feedback} and \texttt{suggested\_fix}.
\emph{Hallucination proxy} is $1-\text{grounding}$, where grounding is a token-overlap heuristic: the fraction of output tokens that also occur in the UI snapshot text (higher grounding suggests less invention under a fixed snapshot).
The tokenizer applies a stop list that excludes short tokens and common UI terms (e.g., ``button'', ``menu'', ``screen''); see Section~\ref{sec:limitations} for the implication on reported values.
\emph{Fabrication proxy rate} is the fraction of samples with grounding below a fixed threshold ($\tau=0.25$), where grounding is a token-overlap heuristic against the UI snapshot.
\emph{Q/Cost} is \emph{Quality} divided by \emph{Calls}.
\IfFileExists{agent_ablation_conditions.tex}{\begin{table}[t]
\centering
\small
\caption{Experimental conditions for agent ablations: each row is uniquely identified by the (Suite, Model, Agent, $N$, CandT, Judge, Fallback) tuple; the legacy compound condition label that encoded the same information as a single underscore-separated token is omitted for readability.}
\label{tab:agent-ablation-conditions}
\resizebox{\linewidth}{!}{%
\begin{tabular}{lllrlllrr}
\toprule
Suite & Model & Agent & $N$ & CandT & Judge & Fallback & Calls & Quality \\
\midrule
HQ & gpt-4.1 & Single-pass & 1 & - & - & - & 1.8 & 0.545 \\
HQ & gpt-4.1-mini & Single-pass & 1 & - & - & - & 2 & 0.418 \\
HQ & gpt-5.2 & Single-pass & 1 & - & - & - & 2 & 0.4 \\
HQ & gpt-4.1 & Hybrid & 2 & 0.3 & LLM & Best-of-N & 4.717 & 0.707 \\
HQ & gpt-4.1 & Hybrid & 2 & 0.3 & Auto & Score-then-select & 4.017 & 0.689 \\
HQ & gpt-4.1-mini & Hybrid & 2 & 0.3 & LLM & Best-of-N & 5.85 & 0.455 \\
HQ & gpt-4.1-mini & Hybrid & 2 & 0.3 & Auto & Score-then-select & 5.7 & 0.473 \\
HQ & gpt-5.2 & Hybrid & 2 & 0.3 & LLM & Best-of-N & 6 & 0.4 \\
HQ & gpt-5.2 & Hybrid & 2 & 0.3 & Auto & Score-then-select & 6 & 0.4 \\
HQ & gpt-4.1 & Best-of-N & 2 & 0.3 & LLM & - & 4.067 & 0.608 \\
HQ & gpt-4.1 & Best-of-N & 2 & 0.3 & Auto & - & 3.717 & 0.643 \\
HQ & gpt-4.1-mini & Best-of-N & 2 & 0.3 & LLM & - & 4.033 & 0.455 \\
HQ & gpt-4.1-mini & Best-of-N & 2 & 0.3 & Auto & - & 3.983 & 0.455 \\
HQ & gpt-5.2 & Best-of-N & 2 & 0.3 & LLM & - & 4 & 0.4 \\
HQ & gpt-5.2 & Best-of-N & 2 & 0.3 & Auto & - & 4 & 0.4 \\
\midrule
CQ & gpt-4.1 & Single-pass & 1 & - & - & - & 1.883 & 0.526 \\
CQ & gpt-4.1-mini & Single-pass & 1 & - & - & - & 1.983 & 0.418 \\
CQ & gpt-5.2 & Single-pass & 1 & - & - & - & 2 & 0.4 \\
CQ & gpt-4.1 & Hybrid & 2 & 0.7 & LLM & Best-of-N & 5.15 & 0.726 \\
CQ & gpt-4.1 & Hybrid & 2 & 0.7 & Auto & Score-then-select & 4.483 & 0.725 \\
CQ & gpt-4.1-mini & Hybrid & 2 & 0.7 & LLM & Best-of-N & 5.95 & 0.546 \\
CQ & gpt-4.1-mini & Hybrid & 2 & 0.7 & Auto & Score-then-select & 5.667 & 0.5 \\
CQ & gpt-5.2 & Hybrid & 2 & 0.7 & LLM & Best-of-N & 6 & 0.4 \\
CQ & gpt-5.2 & Hybrid & 2 & 0.7 & Auto & Score-then-select & 6 & 0.4 \\
CQ & gpt-4.1 & Best-of-N & 2 & 0.7 & LLM & - & 4.183 & 0.708 \\
CQ & gpt-4.1 & Best-of-N & 2 & 0.7 & Auto & - & 3.533 & 0.726 \\
CQ & gpt-4.1-mini & Best-of-N & 2 & 0.7 & LLM & - & 4.05 & 0.527 \\
CQ & gpt-4.1-mini & Best-of-N & 2 & 0.7 & Auto & - & 3.883 & 0.503 \\
CQ & gpt-5.2 & Best-of-N & 2 & 0.7 & LLM & - & 4 & 0.4 \\
CQ & gpt-5.2 & Best-of-N & 2 & 0.7 & Auto & - & 4 & 0.4 \\
\bottomrule
\end{tabular}
}
\end{table}
}{}
\IfFileExists{agent_ablation_hq.tex}{\begin{table}[t]
\centering
\small
\caption{Agent ablation (hallucination-quality focus).}
\label{tab:agent-ablation-hq}
\resizebox{0.98\linewidth}{!}{%
\begin{tabular}{llllrrrrrrr}
\toprule
Model & Agent & Judge & Fallback & $N$ & Prefix ok & Calls & Quality & Q/Cost & Hall.\ proxy & Fab.\ proxy \\
\midrule
gpt-4.1 & Single-pass & - & - & 1 & 0.267 & 1.8 & 0.545 & 0.303 & 0.965 & 0.983 \\
gpt-4.1 & Hybrid & Auto & Score-then-select & 2 & 0.533 & 4.017 & 0.689 & 0.172 & 0.93 & 0.983 \\
gpt-4.1 & Hybrid & LLM & Best-of-N & 2 & 0.567 & 4.717 & 0.707 & 0.15 & 0.938 & 1 \\
gpt-4.1 & Best-of-N & Auto & - & 2 & 0.45 & 3.717 & 0.643 & 0.173 & 0.937 & 0.983 \\
gpt-4.1 & Best-of-N & LLM & - & 2 & 0.383 & 4.067 & 0.608 & 0.149 & 0.944 & 0.95 \\
gpt-4.1-mini & Single-pass & - & - & 1 & 0.033 & 2 & 0.418 & 0.209 & 0.996 & 1 \\
gpt-4.1-mini & Hybrid & Auto & Score-then-select & 2 & 0.133 & 5.7 & 0.473 & 0.083 & 0.982 & 1 \\
gpt-4.1-mini & Hybrid & LLM & Best-of-N & 2 & 0.1 & 5.85 & 0.455 & 0.078 & 0.985 & 1 \\
gpt-4.1-mini & Best-of-N & Auto & - & 2 & 0.1 & 3.983 & 0.455 & 0.114 & 0.989 & 1 \\
gpt-4.1-mini & Best-of-N & LLM & - & 2 & 0.1 & 4.033 & 0.455 & 0.113 & 0.986 & 1 \\
gpt-5.2 & Single-pass & - & - & 1 & 0 & 2 & 0.4 & 0.2 & 1 & 1 \\
gpt-5.2 & Hybrid & Auto & Score-then-select & 2 & 0 & 6 & 0.4 & 0.067 & 1 & 1 \\
gpt-5.2 & Hybrid & LLM & Best-of-N & 2 & 0 & 6 & 0.4 & 0.067 & 1 & 1 \\
gpt-5.2 & Best-of-N & Auto & - & 2 & 0 & 4 & 0.4 & 0.1 & 1 & 1 \\
gpt-5.2 & Best-of-N & LLM & - & 2 & 0 & 4 & 0.4 & 0.1 & 1 & 1 \\
\bottomrule
\end{tabular}
}
\end{table}
}{}
\IfFileExists{agent_ablation_cq.tex}{\begin{table}[t]
\centering
\small
\caption{Agent ablation (cost-quality focus).}
\label{tab:agent-ablation-cq}
\resizebox{0.98\linewidth}{!}{%
\begin{tabular}{llllrrrrrrr}
\toprule
Model & Agent & Judge & Fallback & $N$ & Prefix ok & Calls & Quality & Q/Cost & Hall.\ proxy & Fab.\ proxy \\
\midrule
gpt-4.1 & Single-pass & - & - & 1 & 0.233 & 1.883 & 0.526 & 0.28 & 0.963 & 0.983 \\
gpt-4.1 & Hybrid & Auto & Score-then-select & 2 & 0.6 & 4.483 & 0.725 & 0.162 & 0.925 & 0.983 \\
gpt-4.1 & Hybrid & LLM & Best-of-N & 2 & 0.6 & 5.15 & 0.726 & 0.141 & 0.93 & 0.983 \\
gpt-4.1 & Best-of-N & Auto & - & 2 & 0.6 & 3.533 & 0.726 & 0.205 & 0.928 & 1 \\
gpt-4.1 & Best-of-N & LLM & - & 2 & 0.567 & 4.183 & 0.708 & 0.169 & 0.936 & 1 \\
gpt-4.1-mini & Single-pass & - & - & 1 & 0.033 & 1.983 & 0.418 & 0.211 & 0.995 & 1 \\
gpt-4.1-mini & Hybrid & Auto & Score-then-select & 2 & 0.183 & 5.667 & 0.5 & 0.088 & 0.975 & 0.983 \\
gpt-4.1-mini & Hybrid & LLM & Best-of-N & 2 & 0.267 & 5.95 & 0.546 & 0.092 & 0.961 & 1 \\
gpt-4.1-mini & Best-of-N & Auto & - & 2 & 0.2 & 3.883 & 0.503 & 0.13 & 0.975 & 1 \\
gpt-4.1-mini & Best-of-N & LLM & - & 2 & 0.233 & 4.05 & 0.527 & 0.13 & 0.966 & 0.967 \\
gpt-5.2 & Single-pass & - & - & 1 & 0 & 2 & 0.4 & 0.2 & 1 & 1 \\
gpt-5.2 & Hybrid & Auto & Score-then-select & 2 & 0 & 6 & 0.4 & 0.067 & 1 & 1 \\
gpt-5.2 & Hybrid & LLM & Best-of-N & 2 & 0 & 6 & 0.4 & 0.067 & 1 & 1 \\
gpt-5.2 & Best-of-N & Auto & - & 2 & 0 & 4 & 0.4 & 0.1 & 1 & 1 \\
gpt-5.2 & Best-of-N & LLM & - & 2 & 0 & 4 & 0.4 & 0.1 & 1 & 1 \\
\bottomrule
\end{tabular}
}
\end{table}
}{}
\IfFileExists{agent_output_examples.tex}{\begin{table}[t]
\centering
\small
\caption{Example outputs for the same (persona, journey, task) under different agent strategies (HQ suite; model=gpt-4.1; $n=60$).}
\label{tab:agent-output-examples}
\resizebox{0.98\linewidth}{!}{%
\begin{tabular}{lp{0.46\linewidth}p{0.46\linewidth}}
\toprule
Setting & Friction point & Suggested fix \\
\midrule
Single-pass & navigation: Sulit menemukan slider threshold dengan cepat & Buat slider threshold lebih jelas atau tambahkan highlight. \\
Best-of-2 (LLM judge) & navigation: Threshold sliders are not clearly labeled for each gate & Add clear labels above each threshold slider. \\
Best-of-2 (auto) & navigation: Sulit menemukan slider threshold dengan cepat & Buat slider threshold lebih menonjol atau beri highlight. \\
\bottomrule
\end{tabular}
}
\end{table}
}{}
\UXResolve{\uxparetopdf}{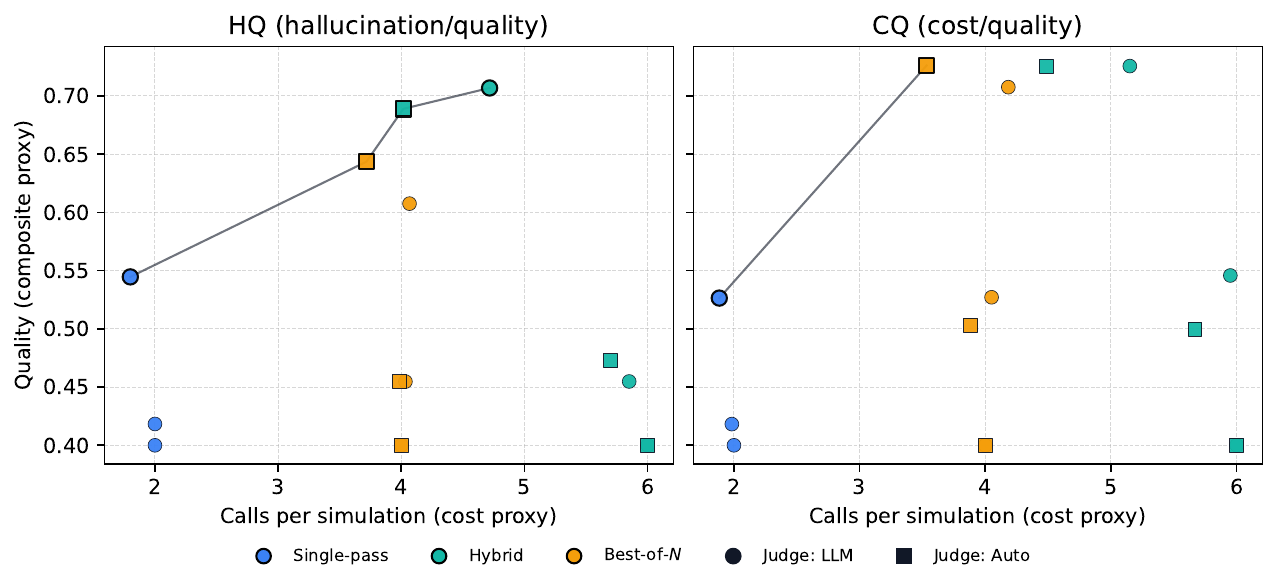}
\ifx\uxparetopdf\empty\else
\begin{figure}[t]
  \centering
  \includegraphics[width=0.98\linewidth]{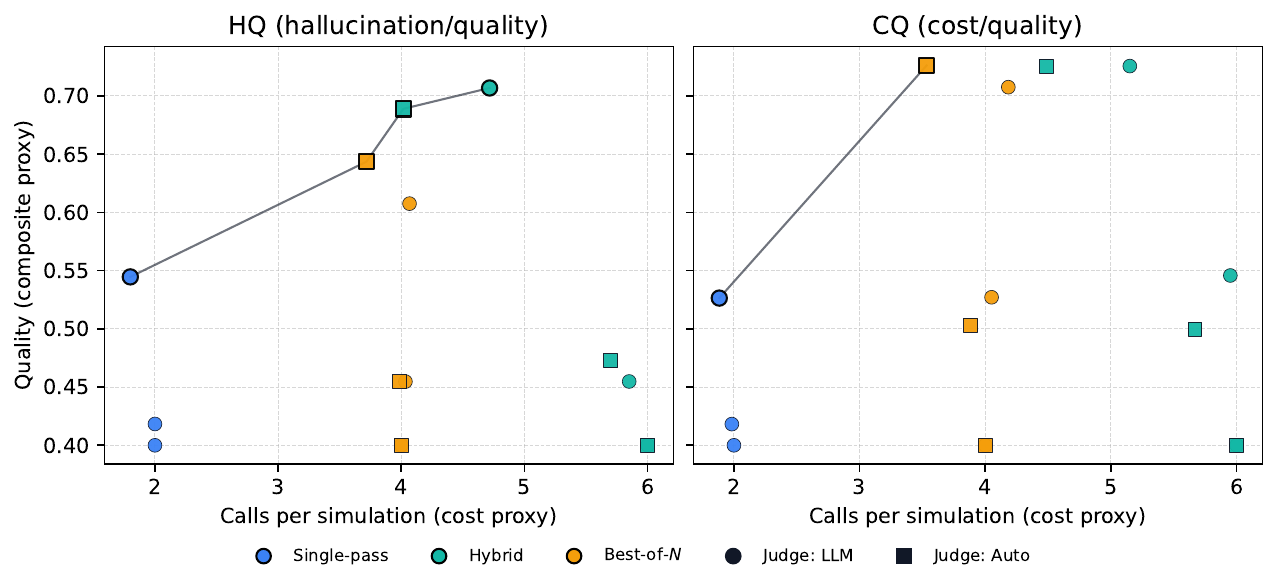}
  \caption{Cost--quality tradeoff for agent strategies. The Pareto frontier (highlighted) is the set of strategies that no other strategy beats on both cost and quality at once. Markers encode judge mode: circle=LLM judge, square=auto-selection.}
  \label{fig:agent-ablation-pareto}
\end{figure}
\fi

\subsection{Score-then-select judge: experimental setup}
\label{sec:anchoring-aware-experiment}
We instantiate the proposed score-then-select judge (Section~\ref{sec:anchoring-aware-judge}) with the same $N$, candidate temperature, and seed settings as best-of-$N$, swapping only the selection protocol. In the agent ablation tables we report $N$ explicitly: for hybrid, $N$ denotes the \emph{maximum} candidate pool used on fallback, and the effective average cost is captured by the \emph{Calls} column. For runtime and cost control across models, we run best-of-$N$ and score-then-select with $N=2$. Prompt excerpts for the simulator and both judges are provided in Appendix~(\emph{Prompt Excerpts}).

\subsection{Proxy alignment (lexical vs TF-IDF vs embeddings)}
We compute friction counts for each proxy dataset under each alignment method, then compute
top-$k$ Jaccard and weighted-Jaccard against the simulation friction distribution.
Unless otherwise stated, we use $k=8$, TF-IDF threshold $t=0.05$, and embedding threshold $t=0.35$ with a fixed embedding subsample of $n=200$ texts.
\emph{Compute note:} for embedding alignment on the larger app-review proxies (\DatasetTinder{} and \DatasetGojek{}), we pre-cap the corpus at $n=10{,}000$ texts before the BGE-M3 subsample of $n=200$ kicks in; on \DatasetTwitter{} and \DatasetAmazon{} the pre-cap is not applied because the per-corpus subsample alone bounds embedding cost.
For the LLM label evaluation on the annotated Amazon subset, we run a fixed subsample of $n=500$ for cost control (Table~\ref{tab:run-contract}).
\IfFileExists{run_contract.tex}{\begin{table}[t]
\centering
\small
\caption{Protocol summary (defaults).}
\label{tab:run-contract}
\begin{tabular}{ll}
\toprule
Setting & Value \\
\midrule
Taxonomy & friction\_taxonomy\_v1 (fixed) \\
Top-$k$ for $J_k$ & 8 \\
TF-IDF threshold & 0.05 \\
Embedding model & BGE-M3 \\
Embedding threshold & 0.35 \\
Embedding subsample per corpus & 200 \\
LLM label eval model & gpt-4.1 \\
LLM label eval sample ($n$) & 500 \\
\bottomrule
\end{tabular}
\end{table}
}{}

\subsection{Sensitivity analysis}
We vary top-$k$ (lexical) and similarity thresholds (TF-IDF and embeddings) to test stability and to
avoid trivial overlap artifacts (e.g., $k$ too large).

\subsection{Bootstrap confidence intervals}
We estimate bootstrap confidence intervals for alignment metrics across multiple method--dataset pairs (lexical, TF-IDF, and embedding on the Amazon, Tinder, and Gojek proxies) to quantify variance and to test whether the single-shot point estimates of $W$ are stable under resampling.

\subsection{LLM label evaluation}
\label{sec:appstore-eval}
We evaluate the LLM's ability to map raw review text to a fixed appstore issue taxonomy using the annotated Amazon subset as a noisy gold label. To keep the policy of the main UX simulator track consistent, this evaluation uses the same Azure-hosted provider family (\texttt{gpt-4.1}) under a single-pass, structured-output classification protocol on a fixed $n=500$ subsample. We report accuracy and macro-F1 (per-category F1 averaged with equal weight, so frequent categories do not dominate the score) against the dataset's category labels, and contrast the LLM result against cheap baselines (majority class, lexical taxonomy match, TF-IDF with logistic regression, and embeddings with logistic regression) computed on the same subsample.

\subsection{Artifacts and reproducibility}
All runs emit artifact-first outputs (CSV/JSON/LaTeX), enabling traceable iteration and paper-grade reporting.
In particular, every table and figure in this draft is exported automatically from versioned run artifacts by a build script, so the manuscript stays in sync with the underlying experiments.

\section{Results}
\subsection{Alignment results}
\IfFileExists{alignment_main.tex}{\begin{table}[t]
\centering
\small
\caption{Proxy alignment summary (top-$k$ Jaccard and weighted-Jaccard). For embedding alignment, we embed a subsample of $n=200$ texts for local compute; on large app-review corpora, we additionally cap the proxy slice at $n=10{,}000$ texts.}
\label{tab:alignment-main}
\begin{tabular}{llrrr}
\toprule
Proxy & Method & $k$ & $J_k$ & $W$ \\
\midrule
\DatasetTinder & \MethodLex & 8 & 0.6 & 0.001 \\
\DatasetTinder & \MethodEmbed & 8 & 0.6 & 0.087 \\
\DatasetGojek & \MethodLex & 8 & 0.6 & 0 \\
\DatasetGojek & \MethodEmbed & 8 & 1 & 0.128 \\
\DatasetTwitter & \MethodLex & 8 & 0.6 & 0 \\
\DatasetTwitter & \MethodEmbed & 8 & 0.778 & 0.123 \\
\DatasetAmazon & \MethodLex & 8 & 0.6 & 0.009 \\
\DatasetAmazon & \MethodTFIDF & 8 & 0.6 & 0.001 \\
\DatasetAmazon & \MethodEmbed & 8 & 0.778 & 0.119 \\
\DatasetAmazonAnn & \MethodLex & 8 & 0.6 & 0.013 \\
\DatasetAmazonAnn & \MethodTFIDF & 8 & 0.778 & 0.003 \\
\DatasetAmazonAnn & \MethodEmbed & 8 & 0.778 & 0.061 \\
\bottomrule
\end{tabular}
\end{table}
}{}
\UXResolve{\uxalignpdf}{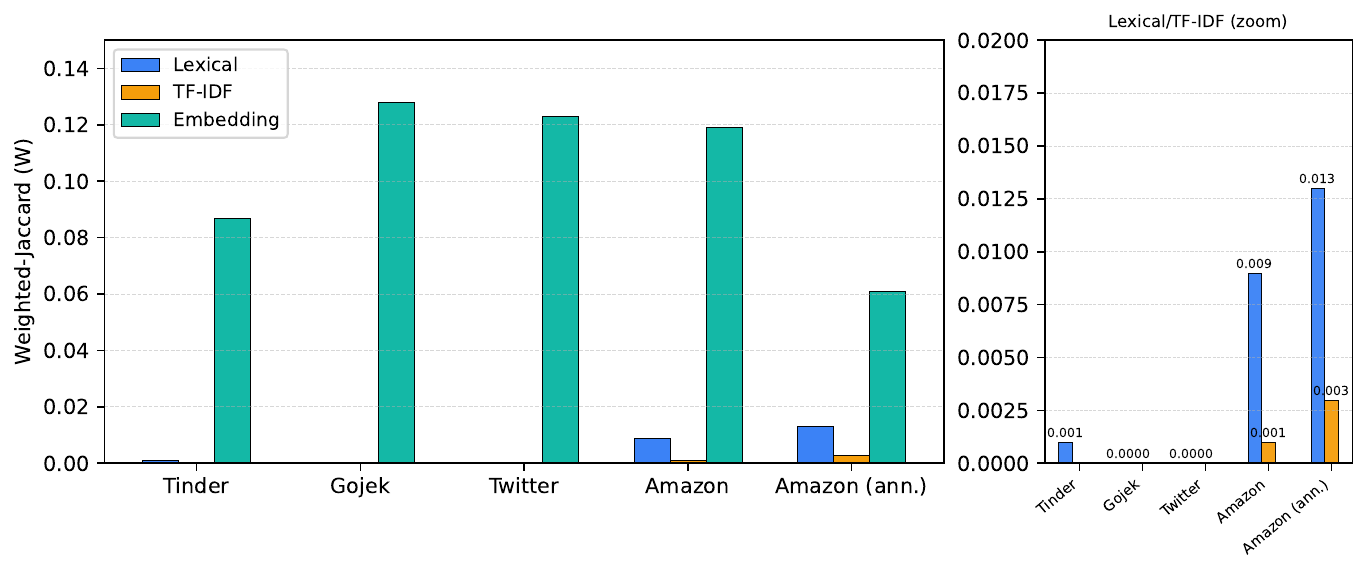}
\ifx\uxalignpdf\empty
  \UXResolve{\uxaligntex}{figures/alignment_plot.tex}
  \ifx\uxaligntex\empty\else
  \begin{figure}[t]
    \centering
    \resizebox{0.98\linewidth}{!}{\begin{tikzpicture}
\begin{axis}[
width=0.98\linewidth,
height=0.52\linewidth,
ybar,
bar width=8pt,
ylabel={Weighted-Jaccard ($W$)},
symbolic x coords={tinder,gojek,twitter,amazon,amazon-ann},
xtick=data,
xticklabels={Tinder,Gojek,Twitter,Amazon,Amazon (ann.)},
ymin=0,
ymax=0.15,
grid=major,
legend style={font=\small, at={(0.02,0.98)}, anchor=north west},
tick label style={font=\small},
label style={font=\small},
]
\addplot[fill=blue!60, draw=blue!80] coordinates {(tinder,0.0010) (gojek,0.0000) (twitter,0.0000) (amazon,0.0090) (amazon-ann,0.0130)};
\addlegendentry{\MethodLex}
\addplot[fill=orange!60, draw=orange!80] coordinates {(amazon,0.0010) (amazon-ann,0.0030)};
\addlegendentry{\MethodTFIDF}
\addplot[fill=teal!60, draw=teal!80] coordinates {(tinder,0.0870) (gojek,0.1280) (twitter,0.1230) (amazon,0.1190) (amazon-ann,0.0610)};
\addlegendentry{\MethodEmbed}
\end{axis}
\end{tikzpicture}}
    \caption{Proxy alignment by dataset and method (weighted-Jaccard; higher is better).}
    \label{fig:alignment}
  \end{figure}
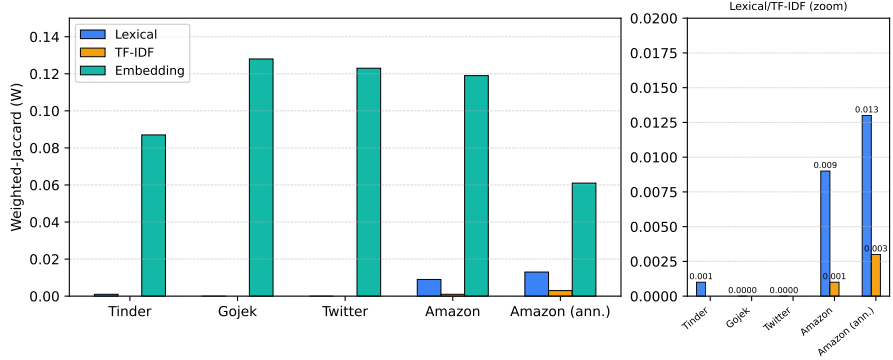
  \fi
\else
  \begin{figure}[t]
    \centering
    \includegraphics[width=0.98\linewidth]{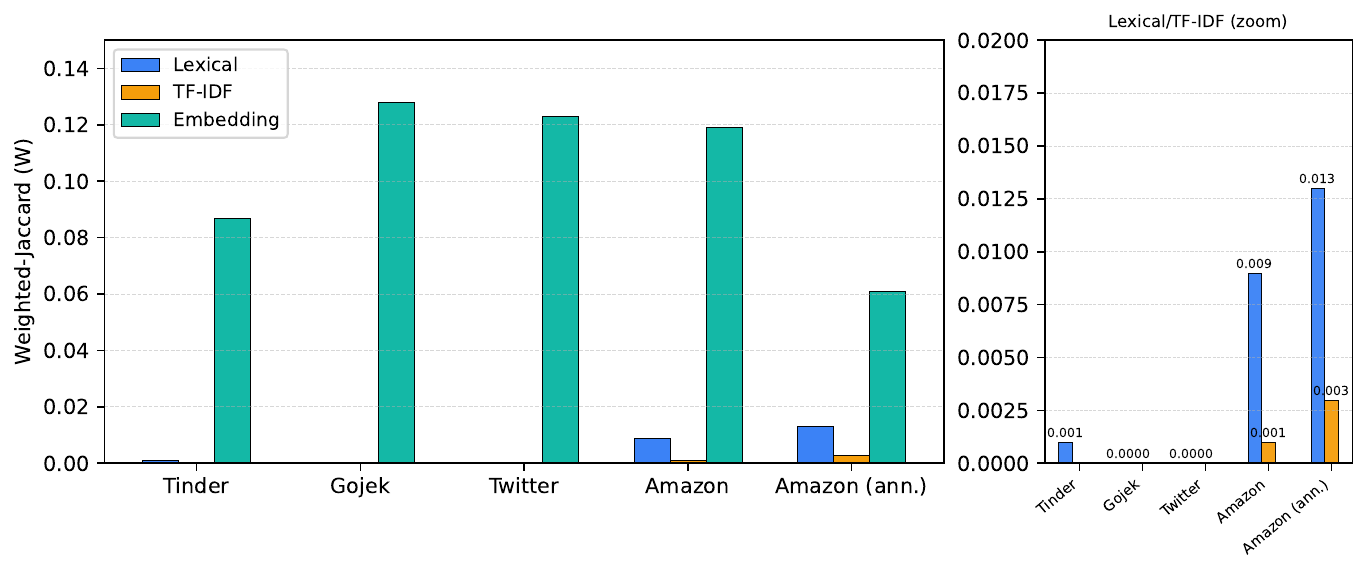}
    \caption{Proxy alignment by dataset and method (weighted-Jaccard; higher is better).}
    \label{fig:alignment}
  \end{figure}
\fi

Across our main proxy corpora (app reviews and support tweets), \MethodEmbed{} produces higher weighted-Jaccard overlap than lexical or TF-IDF baselines (Figure~\ref{fig:alignment}), suggesting that
multilingual semantic similarity is better aligned with our friction taxonomy than keyword matching.
However, absolute overlaps remain modest (Table~\ref{tab:alignment-main}), which is expected: proxy corpora are noisy mixtures of domains and user populations,
while simulation outputs are constrained by our prompt and taxonomy.
In OSS issue proxies, we observe exceptions where lexical overlap can exceed embedding overlap (see discussion below), so we treat the embedding advantage as a tendency rather than a universal guarantee.

\paragraph{Interpreting overlap (example).}
\UXResolve{\uxdistpdf}{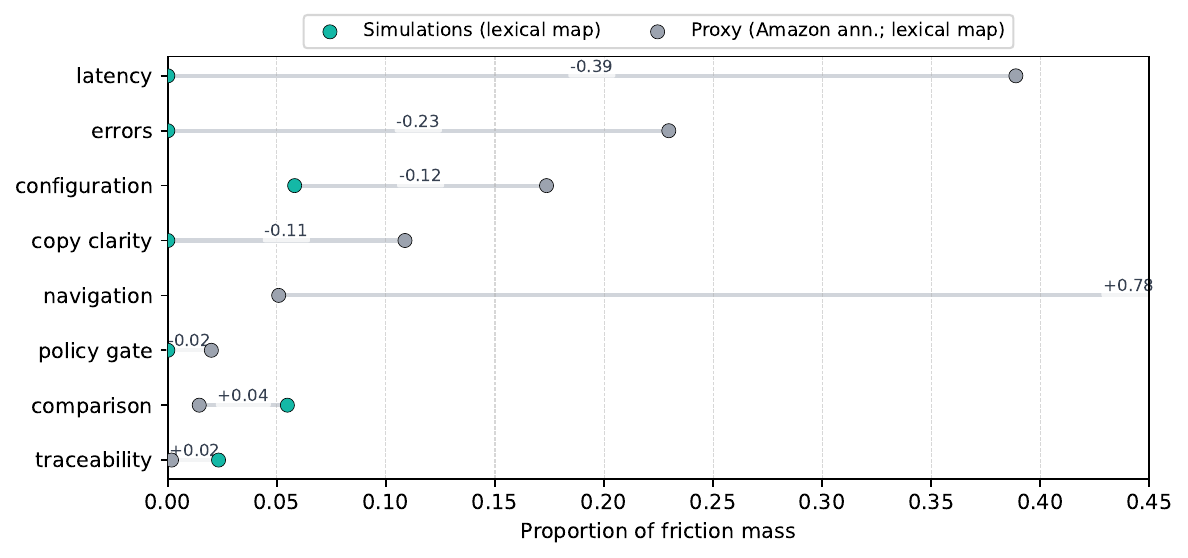}
\ifx\uxdistpdf\empty
  \UXResolve{\uxdisttex}{figures/distribution_amazon_ann.tex}
  \ifx\uxdisttex\empty\else
  \begin{figure}[t]
    \centering
    \begin{tikzpicture}
\begin{axis}[
ybar,
bar width=7pt,
width=0.98\linewidth,
height=0.42\linewidth,
ylabel={Proportion of friction mass},
symbolic x coords={navigation,latency,errors,configuration,copy_clarity,comparison,traceability,policy_gate},
xtick=data,
xticklabels={navigation,latency,errors,configuration,copy clarity,comparison,traceability,policy gate},
xticklabel style={font=\scriptsize, rotate=25, anchor=east},
tick label style={font=\small},
label style={font=\small},
ymin=0,
ymax=1.0,
grid=major,
legend style={font=\small, at={(0.02,0.98)}, anchor=north west},
]
\addplot[fill=teal!55, draw=teal!80] coordinates {(navigation,0.8306) (latency,0.0000) (errors,0.0000) (configuration,0.0581) (copy_clarity,0.0000) (comparison,0.0548) (traceability,0.0233) (policy_gate,0.0000)};
\addlegendentry{Simulations (lexical map)}
\addplot[fill=gray!35, draw=gray!60] coordinates {(navigation,0.0508) (latency,0.3888) (errors,0.2297) (configuration,0.1736) (copy_clarity,0.1087) (comparison,0.0144) (traceability,0.0017) (policy_gate,0.0200)};
\addlegendentry{Proxy (Amazon ann.; lexical map)}
\end{axis}
\end{tikzpicture}
    \caption{Friction-theme distributions for a representative gpt-4.1 \texttt{v2} simulation run vs.\ the Amazon-annotated proxy corpus (lexical mapping; top 8 categories). The distributions diverge sharply: simulated friction concentrates almost entirely on \emph{navigation}---every friction point in this run is emitted with an explicit \texttt{navigation:} taxonomy prefix---whereas the proxy corpus is dominated by \emph{latency} and \emph{errors}. The figure diagnoses \emph{where} simulations and proxies diverge rather than serving as a target to match, and the prefix-driven concentration shows that lexical mapping of taxonomy-prefixed output is sensitive to the emitted prefix, motivating the embedding baseline.}
    \label{fig:distribution-example}
  \end{figure}
  \fi
\else
  \begin{figure}[t]
    \centering
    \includegraphics[width=0.98\linewidth]{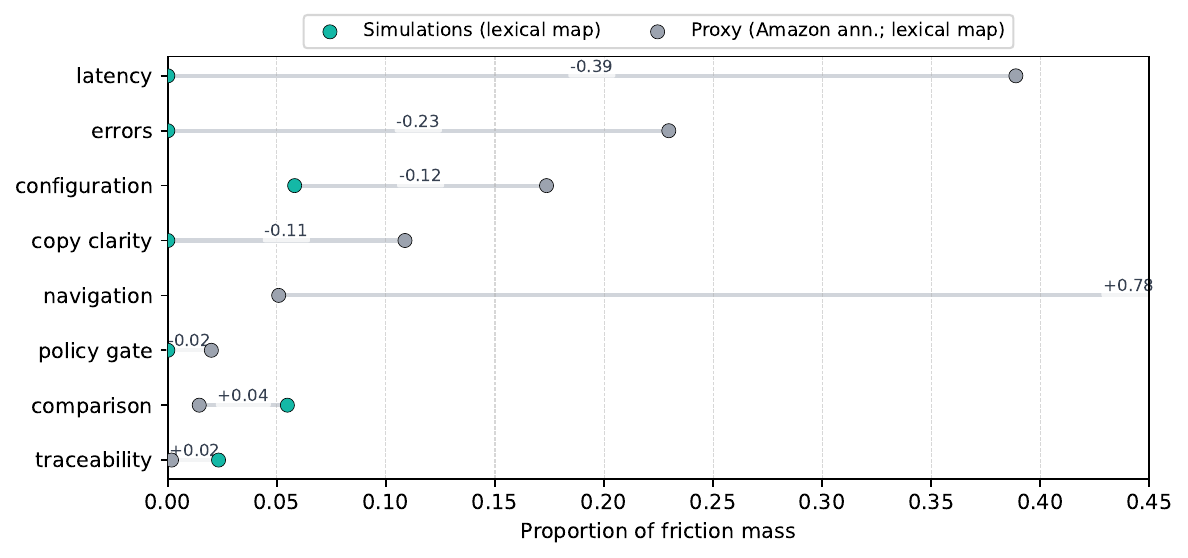}
    \caption{Friction-theme distributions for a representative gpt-4.1 \texttt{v2} simulation run vs.\ the Amazon-annotated proxy corpus (lexical mapping; top 8 categories). The distributions diverge sharply: simulated friction concentrates almost entirely on \emph{navigation}---every friction point in this run is emitted with an explicit \texttt{navigation:} taxonomy prefix---whereas the proxy corpus is dominated by \emph{latency} and \emph{errors}. The figure diagnoses \emph{where} simulations and proxies diverge rather than serving as a target to match, and the prefix-driven concentration shows that lexical mapping of taxonomy-prefixed output is sensitive to the emitted prefix, motivating the embedding baseline.}
    \label{fig:distribution-example}
  \end{figure}
\fi
The goal is not to achieve perfect distributional match; rather, Figure~\ref{fig:distribution-example} helps diagnose \emph{where} simulations and proxies diverge.
Note that the aggregate $W$ values in Table~\ref{tab:alignment-main} are computed from category-level counts accumulated across the full proxy corpus under a shared taxonomy, and may differ from the per-run distribution bars shown here.
The sensitivity rows in Table~\ref{tab:alignment-sensitivity} were executed on a separate simulation pool dedicated to sweeping $k$ and threshold settings, which is why their absolute $W$ values for the same proxy/method pair (e.g., \DatasetAmazonAnn{} lexical, $W=0.017$) can differ slightly from the canonical entry in Table~\ref{tab:alignment-main} ($W=0.013$); we treat Table~\ref{tab:alignment-main} as the canonical reference and use the sensitivity rows only for relative comparisons across $k$ and thresholds.

\paragraph{Key takeaways.}
\begin{itemize}
  \item \MethodEmbed{} typically yields higher $W$ than \MethodLex{} on our primary proxies, making it a strong default under a fixed taxonomy.
  \item $J_k$ is useful for quick inspection but can mislead under thresholding or large $k$; we treat $W$ as the primary signal.
  \item The goal is not to ``prove UX quality'' but to enable early-stage iteration with auditable artifacts and bounded claims.
\end{itemize}

\paragraph{What a consumer can do with these results.}
In practice, the main value is comparative and diagnostic: a team can compare prompt/taxonomy versions (or model settings)
and see whether friction distributions move in the expected direction across proxies, rather than relying on anecdotes.
The same artifacts also support a pragmatic workflow: pick the top-$k$ friction themes to tackle, ship a change, and rerun the pipeline to check
whether the proxy-aligned distribution shifts (and whether the shift is stable under bootstrap intervals).

\paragraph{B2B proxy (OSS issues).}
On GitHub issues from observability repositories, embedding-based alignment tends to outperform lexical matching (Table~\ref{tab:alignment-oss}).
\IfFileExists{alignment_oss.tex}{\begin{table}[t]
\centering
\small
\caption{OSS issues proxy alignment (selected repositories).}
\label{tab:alignment-oss}
\begin{tabular}{lrr}
\toprule
Repository & $W$ (\MethodLex) & $W$ (\MethodEmbed) \\
\midrule
Grafana & 0.12 & 0.185 \\
Jaeger & 0.078 & 0.1 \\
OpenTelemetry Collector & 0.1 & 0.157 \\
OpenTelemetry JS & 0.101 & 0.103 \\
Prometheus & 0.199 & 0.254 \\
Sentry & 0.138 & 0.072 \\
Zipkin & 0.073 & 0.208 \\
\bottomrule
\end{tabular}
\end{table}
}{}
While these proxies are not controlled UX studies, they broaden the evidence beyond app reviews and help stress-test whether our taxonomy and prompts
generalize to B2B tooling discussions.
We also observe failure modes: for some repositories (e.g., Sentry in our selection), lexical overlap can exceed embedding overlap,
suggesting that issue text may be dominated by product-specific vocabulary that semantic embeddings smooth over.

We also observe that top-$k$ Jaccard can be misleading when $k$ is too large (Table~\ref{tab:alignment-sensitivity}).
In particular, lexical $J_k$ reaches 1.0 at $k=10$ on the Amazon annotated proxy while weighted-Jaccard remains unchanged,
supporting our choice to emphasize $W$ for distributional comparisons and to keep $k$ smaller than the taxonomy size.

\subsection{Bootstrap confidence intervals}
\label{sec:results-bootstrap}
\IfFileExists{bootstrap_ci.tex}{\begin{table}[t]
\centering
\small
\caption{Bootstrap confidence intervals for alignment metrics (Amazon low-rated; lexical).}
\label{tab:bootstrap-ci}
\begin{tabular}{lrr}
\toprule
Metric & Mean & 95\% CI \\
\midrule
$J_k$ & 0.601 & [0.6, 0.6] \\
$W$ & 0.011 & [0.01, 0.011] \\
\bottomrule
\end{tabular}
\end{table}
}{}
\UXResolve{\uxbootpdf}{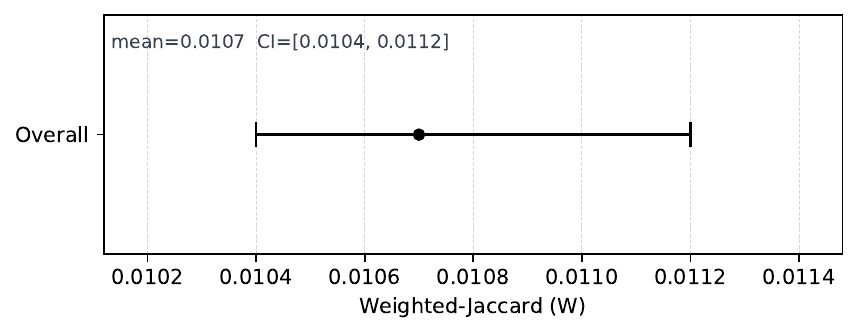}
\ifx\uxbootpdf\empty
  \UXResolve{\uxboottex}{figures/bootstrap_ci_plot.tex}
  \ifx\uxboottex\empty\else
  \begin{figure}[t]
    \centering
    \begin{tikzpicture}
\begin{axis}[
width=0.62\linewidth,
height=0.32\linewidth,
ylabel={Weighted-Jaccard ($W$)},
xtick=\empty,
xmin=-0.5, xmax=0.5,
grid=major,
tick label style={font=\small},
label style={font=\small},
]
\addplot[black, thick] coordinates {(0,0.01040) (0,0.01120)};
\addplot[only marks, mark=*, mark size=1.8pt] coordinates {(0,0.01070)};
\end{axis}
\end{tikzpicture}
    \caption{Bootstrap confidence interval for weighted-Jaccard on Amazon low-rated (lexical).}
    \label{fig:bootstrap-ci}
  \end{figure}
  \fi
\else
  \begin{figure}[t]
    \centering
    \includegraphics[width=0.72\linewidth]{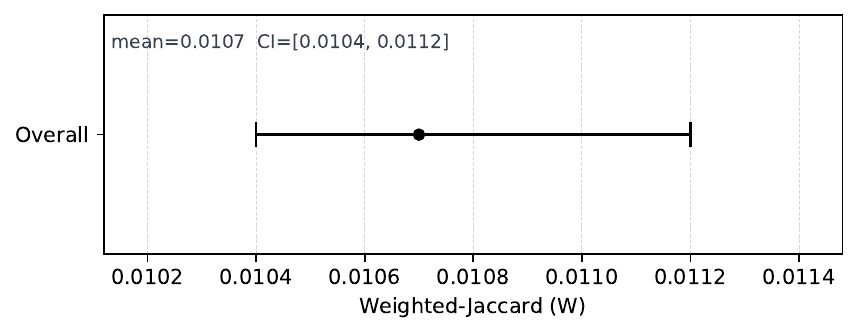}
    \caption{Bootstrap confidence interval for weighted-Jaccard on Amazon low-rated (lexical).}
    \label{fig:bootstrap-ci}
  \end{figure}
\fi
Bootstrap intervals (Table~\ref{tab:bootstrap-ci}; visualized in Figure~\ref{fig:bootstrap-ci}) are tight for the Amazon low-rated proxy under the lexical baseline, suggesting that the aggregate category distribution is stable at this scale for our chosen taxonomy. Table~\ref{tab:bootstrap-ci-multi} extends this to $8$ method--dataset pairs in total: lexical on \DatasetAmazon, \DatasetTinder, and \DatasetGojek; TF-IDF on \DatasetAmazon{} and \DatasetAmazonAnn; and embedding (BGE-M3) on \DatasetAmazon, \DatasetTinder, and \DatasetGojek. Two observations follow: (i)~lexical and TF-IDF baselines on app-review proxies yield $W$ means rounding to $0$ with CIs $[0,0]$, which is "stable at zero" rather than "aligned"---i.e., these baselines deterministically fail to align under bootstrap resampling, consistent with the point estimates in Table~\ref{tab:alignment-main}; (ii)~$J_k$ on Tinder and Gojek shows mild but non-zero variance (CI width $\approx 0.18$) because top-$k$ set membership flickers near the boundary, while $W$ remains numerically zero, reinforcing that $J_k$ overstates alignment when the underlying mass overlap is negligible (the same effect documented in H2). The three embedding configurations are discussed under H3 below.
\IfFileExists{bootstrap_ci_multi.tex}{\begin{table}[t]
\centering
\small
\caption{Bootstrap confidence intervals for alignment metrics across method--dataset pairs (200 iterations for lexical/TF-IDF, 50 for embedding; bootstrap resample size noted as $n$).}
\label{tab:bootstrap-ci-multi}
\resizebox{0.98\linewidth}{!}{%
\begin{tabular}{llrrrrr}
\toprule
Dataset & Method & $n$ & $J_k$ mean & $J_k$ 95\% CI & $W$ mean & $W$ 95\% CI \\
\midrule
\DatasetAmazon & \MethodLex & 79531 & 0.601 & [0.6, 0.6] & 0.0107 & [0.0104, 0.0112] \\
\DatasetAmazon & \MethodTFIDF & 10000 & 0.6 & [0.6, 0.6] & 0 & [0, 0] \\
\DatasetAmazonAnn & \MethodTFIDF & 8971 & 0.6 & [0.6, 0.6] & 0 & [0, 0] \\
\DatasetAmazon & \MethodEmbed & 200 & 0.804 & [0.778, 1] & 0 & [0, 0] \\
\DatasetTinder & \MethodLex & 10000 & 0.764 & [0.6, 0.778] & 0 & [0, 0] \\
\DatasetTinder & \MethodEmbed & 200 & 0.782 & [0.778, 0.778] & 0 & [0, 0] \\
\DatasetGojek & \MethodLex & 10000 & 0.74 & [0.6, 0.778] & 0 & [0, 0] \\
\DatasetGojek & \MethodEmbed & 200 & 0.774 & [0.778, 0.778] & 0 & [0, 0] \\
\bottomrule
\end{tabular}
}
\end{table}
}{}

\subsection{Failure-mode analysis: examples flagged as fabricated}
To make the grounding/fabrication proxies (Section~\ref{sec:limitations}) concrete, Table~\ref{tab:fabrication-examples} lists representative outputs that the adversarial judge flagged as fabricated, including the unsupported span and rationale. We note that the adversarial judge is intentionally one-sided---it emits items only when unsupported spans are detected---so every row is a positive case by construction; non-fabricated counterfactuals are not produced by this pass. This view is complementary to the aggregate hallucination/fabrication rates in Tables~\ref{tab:agent-ablation-hq} and~\ref{tab:agent-ablation-cq}: the aggregate rates are conservative lower bounds (stop-list induced), while the worked examples illustrate which kinds of friction phrasings the proxy actually catches.
\IfFileExists{fabrication_examples.tex}{\begin{table}[t]
\centering
\footnotesize
\setlength{\tabcolsep}{4pt}
\caption{Top examples flagged as fabricated by an adversarial judge (4 representative cases, HQ ablation; the eval only emits items where unsupported spans are detected, so all rows are positive cases by construction).}
\label{tab:fabrication-examples}
\begin{tabular}{p{0.22\linewidth}rp{0.20\linewidth}p{0.42\linewidth}}
\toprule
Setting & Conf. & Friction & Unsupported span / rationale \\
\midrule
Hybrid (best-of-$N$ fallback, LLM judge) & 0.8 & navigation: threshold sliders unclear & span: Thresholds could be more intuitive. | Claims about UI elements (threshold sliders) are not supported by the UI snapshot text. \\
Best-of-$N$ (LLM judge) & 0.8 & navigation: unclear threshold sliders & span: friction-point: navigation: unclear threshold sliders | Claimed friction point and step 1 friction are supported by the UI snapshot, but the suggested fix is not. The UI snapshot shows thresh... \\
Best-of-$N$ (LLM judge) & 0.8 & navigation: falta de informacion sobre el sistema & span: falta de informacion sobre el sistema | The UI snapshot does not contain information about the suggested fixes or friction points. The sentiment analysis is also neutral, which contradicts th... \\
Hybrid (score-then-select, auto judge) & 0.8 & navigation: falta de contexto para entender a politica & span: falta de contexto para entender a politica | Claims about suggested fixes and friction points are not supported by the UI snapshot text. \\
\bottomrule
\end{tabular}
\end{table}
}{}

\subsection{Azure provider agent-ablation results}
Tables~\ref{tab:agent-ablation-hq} and \ref{tab:agent-ablation-cq} report the reduced ablation suites across three OpenAI deployments on Azure (\texttt{gpt-4.1}, \texttt{gpt-4.1-mini}, \texttt{gpt-5.2}).
\begin{itemize}
  \item \emph{Model sensitivity.} Under the same prompt/schema contract, \texttt{gpt-4.1} is strongest on quality (HQ hybrid-LLM quality $0.707$; CQ hybrid-auto and best-of-$N$ auto both at $0.726$), \texttt{gpt-4.1-mini} is consistently lower but still functional, and \texttt{gpt-5.2} collapses to near-degenerate behavior across every condition we ran (prefix rate $0$, grounding $0$, fabrication proxy $1.0$). The \texttt{gpt-5.2} failure is the only reproducible model-level breakdown we observed.
  \item \emph{Transient failure documented and re-measured.} An initial CQ batch for \texttt{gpt-4.1} (candidate temperature $0.7$, LLM judge) produced degenerate outputs (quality $0.4$, prefix $0$) on the hybrid and best-of-$N$ conditions. We re-executed those conditions with the identical configuration and \emph{the same model converged to non-degenerate behavior} (quality $0.71$--$0.73$, prefix $0.57$--$0.60$); the ablation tables report this latest measurement per condition. We therefore attribute the original CQ-\texttt{v2} LLM-judge failures on \texttt{gpt-4.1} to a transient API/scheduling artifact, not a structural condition-specific failure of the prompt contract. The \texttt{gpt-5.2} collapse, in contrast, reproduced across batches and remains a real model-level failure.
\end{itemize}
The cost--quality frontier (Figure~\ref{fig:agent-ablation-pareto}) summarizes the resulting tradeoffs; auto-judge variants tend to dominate LLM-judge variants on cost without sacrificing quality, and the differences across models dwarf the differences across selection protocols. We report all conditions rather than filtering the \texttt{gpt-5.2} runs, because aggregate tables without model and condition labels would hide that breakdown.

\subsection{LLM label evaluation}
\IfFileExists{appstore_eval.tex}{\begin{table}[t]
\centering
\small
\caption{LLM label evaluation on the Amazon Appstore annotated subset, using the same provider model family as the main UX agent-ablation matrix to keep the local-vs-provider policy consistent across tracks.}
\label{tab:appstore-eval}
\begin{tabular}{lrr}
\toprule
Model & Accuracy & Macro-F1 \\
\midrule
gpt-4.1 (n=500) & 0.556 & 0.438 \\
\bottomrule
\end{tabular}
\end{table}
}{}
On the annotated Amazon subset, \texttt{gpt-4.1} (Azure provider, same family as the main UX ablation) achieves moderate accuracy (Table~\ref{tab:appstore-eval}; accuracy $0.556$, macro-F1 $0.438$), consistent with uneven category difficulty and confusable labels (e.g., UX vs functionality). A cheap-baselines comparison (Table~\ref{tab:appstore-baselines}) sharpens the picture: an embedding + logistic-regression baseline (\texttt{nomic-embed-text}) attains higher accuracy ($0.635$) than the LLM, while the LLM still beats majority, lexical, and TF-IDF baselines on macro-F1. Table~\ref{tab:appstore-confusions} reports the most frequent label confusions and shows that the largest single error mode is \texttt{functionality\_features} being mapped to \texttt{ui\_ux}, which dominates the residual error budget.
Together, these results motivate our focus on distribution-level proxy alignment rather than treating LLM category assignments as reliable ground truth at the single-example level.
\IfFileExists{appstore_baselines.tex}{\begin{table}[t]
\centering
\small
\caption{Cheap baselines for the annotated Appstore subset (same $n$ as Table~\ref{tab:appstore-eval}); embedding baseline uses a subset for cost.}
\label{tab:appstore-baselines}
\begin{tabular}{lrr}
\toprule
Baseline & Accuracy & Macro-F1 \\
\midrule
Majority & 0.432 & 0.086 \\
Lexical taxonomy match & 0.424 & 0.366 \\
TF-IDF + logistic regression (CV) & 0.492 & 0.146 \\
Embedding (nomic-embed-text) + logistic regression (CV) (n=200) & 0.635 & 0.336 \\
\bottomrule
\end{tabular}
\end{table}
}{}
\IfFileExists{appstore_confusions.tex}{\begin{table}[t]
\centering
\small
\caption{Most frequent label confusions in LLM classification (Amazon annotated; n=500).}
\label{tab:appstore-confusions}
\resizebox{0.98\linewidth}{!}{%
\begin{tabular}{llr}
\toprule
Gold & Predicted & Count \\
\midrule
\texttt{performance\_stability} & \texttt{functionality\_features} & 30 \\
\texttt{functionality\_features} & \texttt{other} & 27 \\
\texttt{ui\_ux} & \texttt{functionality\_features} & 23 \\
\texttt{compatibility\_device} & \texttt{functionality\_features} & 20 \\
\texttt{support\_responsiveness} & \texttt{functionality\_features} & 17 \\
\texttt{functionality\_features} & \texttt{performance\_stability} & 16 \\
\texttt{functionality\_features} & \texttt{ui\_ux} & 12 \\
\texttt{compatibility\_device} & \texttt{performance\_stability} & 8 \\
\bottomrule
\end{tabular}
}
\end{table}
}{}

\subsection{Hypothesis evaluation}
We close with an explicit mapping from our hypotheses (Section~1) to results.
\textbf{H1} (embedding outperforms lexical under $W$) is \textbf{supported}: embedding yields higher $W$ on all primary app-review and support-tweet proxies
(e.g., Gojek: $W_\text{embed}=0.128$ vs $W_\text{lex}=0.000$; Amazon Low-rated: $W_\text{embed}=0.119$ vs $W_\text{lex}=0.009$).
The exception is Sentry OSS issues ($W_\text{lex}=0.138 > W_\text{embed}=0.072$), consistent with H1's ``tendency rather than universal'' framing.

\textbf{H2} ($J_k$ saturates and overstates alignment at large $k$) is \textbf{supported}: on the Amazon annotated proxy, lexical $J_k$ jumps from $0.333$ at $k=6$ to $1.0$ at $k=10$ while the weighted-Jaccard $W$ stays flat at $0.017$ across the same sensitivity sweep (canonical $W=0.013$ at $k=8$ in Table~\ref{tab:alignment-main}). $J_k$ is therefore misleading once $k$ approaches the taxonomy size.

\textbf{H3} (aggregate estimates are stable enough for iterative decision-making when reported with CIs) is \textbf{partially supported, with a systematic caveat about embedding-based $W$}. The multi-configuration bootstrap (Table~\ref{tab:bootstrap-ci-multi}; 200 iterations for lexical/TF-IDF, 50 for embedding) covers $8$ method--dataset pairs. Lexical and TF-IDF baselines on app-review proxies produce stable aggregate estimates: their $J_k$ CIs are zero-width or near-zero (top-$k$ set membership rarely changes under resampling), and their $W$ CIs are tight, often pinned at zero (deterministic non-alignment). For the lexical baseline on Amazon low-rated specifically, $W=0.0107$ with CI $[0.0104, 0.0112]$, indicating high stability at scale.

The embedding baseline (BGE-M3, 50 bootstrap iterations at $n=200$) tells a consistent and contrary story across all three proxies we ran. Its $J_k$ mean lands in $[0.77, 0.80]$ with tight CIs at $0.778$, but $W$ collapses to $0$ across all 50 iterations on \DatasetAmazon, \DatasetTinder, and \DatasetGojek alike---in sharp contrast to the single-shot point estimates of $W=0.119$, $0.087$, and $0.128$ respectively reported in Table~\ref{tab:alignment-main}. The three-way replication makes it implausible that the gap is a single-dataset anomaly: the embedding $W$ point estimates at this subsample size are \emph{systematically} unstable under bootstrap resampling. Practitioners should therefore report bootstrap CIs alongside any embedding-based alignment number before treating it as a target. In sum, H3 holds for lexical and TF-IDF baselines and for $J_k$ under embedding, but is \emph{contradicted} for embedding-based $W$ at this subsample size; characterizing the dependence on subsample size, anchor construction, and similarity threshold is left as future work.

\section{Limitations and Threats to Validity}
\label{sec:limitations}
Our approach has important limitations and validity threats.
\paragraph{Construct validity.}
Proxy corpora are noisy mixtures of domains and user populations; alignment should not be interpreted as task success, satisfaction, or usability.
Our overlap metrics measure \emph{distributional similarity of friction themes}, not causal impact.
We partially mitigate this by reporting multiple metrics, sensitivity analyses, and confidence intervals, but the construct remains approximate.

\paragraph{Internal validity.}
LLM simulations can suffer from prompt framing bias and ``algorithmic fidelity'' failures~\cite{amirova2024algorithmicfidelity,gao2024scylla}.
Persona prompts may introduce systematic biases and unrealistic behavior~\cite{li2025personacatch,prpa2024syntheticpersonae}, and prior work shows that
LLM-generated ``synthetic survey'' distributions can diverge from human distributions~\cite{bisbee2024perils}.
We therefore treat simulations as \emph{decision support} and emphasize versioned prompts, schema-constrained outputs, and artifact trails for auditability.

\paragraph{Taxonomy coverage of simulations.}
Under the product context studied here, the \texttt{v2} simulator concentrates friction in a single taxonomy category: across the gpt-4.1 runs, essentially every emitted friction point carries the \texttt{navigation:} taxonomy prefix (Figure~\ref{fig:distribution-example}).
This narrows the friction-theme distribution the pipeline can surface and likely reflects both the prompt's taxonomy-prefix instruction and the navigation-heavy product UI; broader coverage would require more diverse personas, tasks, and UI snapshots.
We therefore treat the friction-theme comparison as a divergence diagnostic rather than evidence of calibrated theme prevalence.

\paragraph{External validity.}
Results depend on the chosen taxonomy, locales, and proxy sources.
While adding OSS issues broadens evidence toward B2B tooling, it remains an indirect proxy and may over-represent power users or issue reporters.
Future work should evaluate domain-specific proxies and compare against controlled studies when available.

\paragraph{Evaluation validity.}
The annotated Amazon subset provides only a weak gold label for category classification and may encode annotator bias.
Even when LLMs are used as evaluators or automatic labelers, they may be biased and inconsistent in rubric application~\cite{liu2023geval,zheng2023mtbench,wang2024notfair}.
We use the labeled subset primarily for calibration and error analysis, not as a definitive performance target.

\paragraph{Grounding heuristic calibration.}
The hallucination proxy ($1-\text{grounding}$) and fabrication proxy rate are computed using a token-overlap heuristic that applies a stop list removing short tokens and common UI terms (e.g., ``button'', ``menu'', ``tab'', ``screen'', ``field'').
Because simulation outputs frequently reference these same UI elements using the same vocabulary, the stop list systematically underestimates true token overlap, yielding hallucination proxy values near 1.0 across all conditions.
These values therefore reflect a conservative lower bound on grounding, not evidence that nearly all output tokens are fabricated; teams should recalibrate the stop list for their UI vocabulary before using this proxy as a go/no-go gate.
The proxy remains useful for \emph{relative} comparisons across agent strategies and prompt versions, where the stop-list bias is consistent.

\paragraph{Cost and model choice.}
Our agent ablation uses Azure-hosted OpenAI models (\texttt{gpt-4.1}, \texttt{gpt-4.1-mini}, \texttt{gpt-5.2}) and the LLM label evaluation uses \texttt{gpt-4.1}; the alignment metrics additionally use a local embedding model (BGE-M3 via Ollama). Stronger paid-model calibration of the alignment pipeline at scale, and a broader sweep over embedding model and subsample size, are deferred to future work to control cost and to avoid iterative overfitting.

\section{Conclusion}
We presented a reproducible UX simulation pipeline and proxy-validation protocol
that compares simulated friction signals against multiple public corpora using two complementary alignment metrics (top-$k$ Jaccard and weighted-Jaccard).

\paragraph{Summary of findings.}
Across six proxy datasets (app reviews, support tweets, and OSS issues), embedding-based alignment (BGE-M3) consistently yielded higher weighted-Jaccard than lexical or TF-IDF baselines on app-review and support-tweet proxies (H1 supported; e.g., Gojek: $W=0.128$ vs $0.000$ lexical), with Sentry OSS issues as a notable exception where product-specific vocabulary favors lexical matching.
Top-$k$ Jaccard reached 1.0 at $k=10$ while $W$ remained at 0.017 on the Amazon annotated proxy, confirming that $J_k$ overstates alignment at large $k$ (H2 supported) and motivating $W$ as the primary signal.
Bootstrap intervals cover $8$ method--dataset pairs. Lexical and TF-IDF baselines are stable ($W$ on Amazon low-rated: $[0.010, 0.011]$, others pinned at $0$). Embedding (BGE-M3) bootstraps at $n=200$ collapse $W$ to $0$ on all three proxies we ran (Amazon, Tinder, Gojek), contradicting the single-shot point estimates of $0.119$, $0.087$, and $0.128$ respectively. H3 is therefore partially supported (lexical/TF-IDF and embedding $J_k$ are stable; embedding $W$ at this subsample size is systematically not), and we recommend bootstrap CIs alongside any embedding-based alignment number before it is used as a decision target. Studying the dependence on subsample size, anchor construction, and similarity threshold is left for future work.
In the agent ablation, \texttt{gpt-4.1} with hybrid strategy achieved the highest quality (HQ suite: $0.707$ with LLM judge; CQ suite: $0.726$, tied between hybrid-auto and best-of-$N$-auto after a confirming rerun of the \texttt{v2} LLM-judge conditions whose initial batch had been transiently degenerate), while \texttt{gpt-5.2} collapsed across every condition we ran and reproduced that collapse across batches---illustrating that model-level breakdowns must be separated from transient API artifacts, and that model- and condition-labeled ablation reporting (with reruns when batch-level degeneration is suspected) is essential.

\paragraph{Limitations recap.}
Proxy corpora are noisy and do not measure task success; the grounding heuristic is sensitive to stop-list choices; and although we now report bootstrap CIs across $8$ method--dataset pairs, the embedding configuration is exercised at a single subsample size and a single anchor construction, so the H3 caveat on embedding stability is itself bounded by those choices.
We treat all outputs as directional signals for early-stage iteration, not as claims about individual user behavior.

\paragraph{Future work.}
We plan to:
\begin{itemize}
    \item expand locale and domain matching and add adversarial negative-control proxies;
    \item run final calibration with stronger paid models at scale;
    \item connect proxy alignment to downstream product decisions and release gating;
    \item report bootstrap CIs across alignment methods and datasets.
\end{itemize}

\paragraph{Product perspective.}
From a consumer standpoint, the key deliverable is a repeatable ``UX check'' that can be run before and after changes,
producing a small set of prioritized friction themes, confidence bounds, and traceable artifacts.
This supports better decision-making under uncertainty: teams can focus on the highest-leverage issues, communicate
risk clearly across stakeholders, and avoid shipping changes that introduce new friction patterns without noticing.

\FloatBarrier
\appendix
\section{Additional Tables}
\IfFileExists{alignment_sensitivity.tex}{\begin{table}[H]
\centering
\small
\caption{Sensitivity analysis on the Amazon annotated proxy (selected settings).}
\label{tab:alignment-sensitivity}
\begin{tabular}{lrrrr}
\toprule
Variant & $k$ & Threshold & $J_k$ & $W$ \\
\midrule
\MethodLex & 6 & - & 0.333 & 0.017 \\
\MethodLex & 10 & - & 1.000 & 0.017 \\
\MethodTFIDF & 8 & 0.05 & 0.778 & 0.003 \\
\MethodTFIDF & 8 & 0.2 & 0.778 & 0.002 \\
\MethodEmbed & 8 & 0.25 & 0.778 & 0.061 \\
\MethodEmbed & 8 & 0.45 & 0.778 & 0.049 \\
\bottomrule
\end{tabular}
\end{table}
}{}

\section{Friction Taxonomy}
Table~\ref{tab:friction-taxonomy} lists the full set of friction categories used in this paper, their default weights, and a truncated set of representative keywords. The complete keyword lists (including Portuguese, Spanish, and Indonesian variants used for multilingual proxies) and the domain-specific weight overrides are kept in the versioned taxonomy configuration.
\IfFileExists{friction_taxonomy.tex}{\begin{table}[H]
\centering
\small
\caption{Friction taxonomy (v1): categories, default weights, and representative keywords (truncated to 6 per row; the complete multilingual keyword list and weight overrides are in the versioned taxonomy configuration).}
\label{tab:friction-taxonomy}
\begin{tabular}{lrp{0.58\linewidth}}
\toprule
Category & Default $w$ & Representative keywords \\
\midrule
\texttt{comparison} & 1 & compare, baseline, diff, delta, comparar, comparacao, ... \\
\texttt{configuration} & 1 & threshold, weight, settings, config, configuracao, ajuste, ... \\
\texttt{copy\_clarity} & 1 & confusing, unclear, ambiguous, wording, confuso, nao claro, ... \\
\texttt{errors} & 1 & error, fail, bug, broken, crash, erro, ... \\
\texttt{export} & 1 & export, exportar, csv, download csv, export csv, download report, ... \\
\texttt{latency} & 1 & slow, lag, timeout, loading, wait, lento, ... \\
\texttt{navigation} & 1 & menu, breadcrumb, where to click, sidebar, tab bar, discover, ... \\
\texttt{policy\_gate} & 1 & policy, denied, blocked, restricted, negado, bloqueado, ... \\
\texttt{search} & 1 & filter, query, findability, search bar, search results, buscar, ... \\
\texttt{traceability} & 1 & trace, traceability, audit, audit trail, run log, audit log, ... \\
\bottomrule
\end{tabular}
\end{table}
}{}

\section{Prompt Excerpts}
\noindent
This appendix provides short excerpts of the key prompts used in our UX micro-simulation pipeline.

\subsection*{Simulation prompt (\texttt{v1})}
\begin{codebox}
System: You are a UX research assistant. Return structured JSON only.
Hard rules: ASCII, no PII/secrets, <=12 words per string, exactly 2 steps,
exactly 1 friction point and 1 HEART GSM item, non-empty feedback/fix.
\end{codebox}

\subsection*{Simulation prompt (\texttt{v2}) --- grounding + taxonomy prefix}
\begin{codebox}
Hard rules (additions): Only reference UI elements in the UI snapshot text.
Friction taxonomy: choose exactly 1 category; prefix as "<category_id>: ...".
\end{codebox}

\subsection*{Judge prompt (best-of-$N$ selection)}
\begin{codebox}
System: You are a strict evaluator. You receive multiple candidate JSON outputs.
Select the best candidate. Return JSON: {"selected_index":0,"scores":[...],"rationale":"..."}.
\end{codebox}

\subsection*{Judge prompt (score-then-select)}
\begin{codebox}
System: You are a strict evaluator. You receive ONE candidate JSON output.
Score independently: schema_compliance, ui_grounding, actionability, overall.
Return JSON: {"schema_compliance":...,"ui_grounding":...,"actionability":...,"overall":...,"rationale":"..."}.
\end{codebox}

\subsection*{Appstore label prompt (LLM evaluation)}
\begin{codebox}
System: You are a strict routing classifier for UX friction categories.
Return only valid JSON.
\end{codebox}
\begin{codebox}
User: Given the review text, assign exactly one category from the taxonomy.
Include a short justification.
\end{codebox}

\section*{Acknowledgments and AI Tools Disclosure}
This research leveraged AI-assisted tools to support code development and manuscript preparation, under full human oversight and accountability. The following tools were used:
\begin{itemize}
    \item \textbf{Language models:} OpenAI Codex (GPT-5 family), Anthropic Claude Code (Opus/Sonnet), and Google Gemini CLI were used for code drafting/review and language refinement of author-written text.
    \item \textbf{Web search:} MCP Tavily integration was used to support literature discovery and fact-checking.
\end{itemize}
All scientific arguments, empirical methodology, statistical analysis, research questions, and conclusions were conceived, developed, and validated by the authors. AI assistance was limited to code drafting/review and language refinement; it was not used to generate or alter experimental data, metrics, or conclusions. Authors reviewed and validated all AI-assisted outputs before inclusion. AI systems were not listed as authors. No generative-AI artwork or figures were used in the manuscript.

\paragraph{Data and code availability.}
The replication package is publicly available at \url{https://github.com/alemaiorano/ux-microsim-replication}. It provides the paper sources, the versioned run artifacts, and the export scripts that regenerate every table and figure, so that all reported results can be independently verified. The simulation engine and its prompt and configuration files are not redistributed; representative prompt excerpts are given in the appendix.

{\sloppy\hbadness=2000\emergencystretch=2em\Urlmuskip=0mu plus 2mu\relax
\bibliographystyle{plain}
\bibliography{references}
}

\end{document}